\documentclass[sigconf, nonacm]{acmart}

\usepackage{float}
\usepackage{cmap}				% Mapear caracteres especiais no PDF		
\usepackage[linesnumbered,ruled,vlined]{algorithm2e}
\usepackage{graphicx} % Required for inserting images
\usepackage{listings}
\usepackage{lscape} 
\usepackage{rotating} %rotates the figures, page
\usepackage{color}
\usepackage{hyperref}
\usepackage{tabularx}
\usepackage{tikz}
\usepackage{pgfplots}
\usepackage{soul}
\usepackage{todonotes}
\usepackage{adjustbox}
\usepackage{xspace}
\usepackage{enumerate}
\definecolor{blue}{RGB}{41,5,195}
\definecolor{gray}{rgb}{.4,.4,.4}
\definecolor{gray}{rgb}{.4,.4,.4}
\definecolor{pblue}{rgb}{0.13,0.13,1}
\definecolor{pgreen}{rgb}{0,0.5,0}
\definecolor{pred}{rgb}{0.9,0,0}
\definecolor{pgrey}{rgb}{0.46,0.45,0.48}
\definecolor{lightgray}{rgb}{0.95, 0.95, 0.96}
\definecolor{whitesmoke}{rgb}{0.96, 0.96, 0.96}
\definecolor{javared}{rgb}{0.6,0,0} % for strings
\definecolor{javagreen}{rgb}{0.25,0.5,0.35} % comments
\definecolor{javapurple}{rgb}{0.5,0,0.35} % keywords
\definecolor{javadocblue}{rgb}{0.25,0.35,0.75} % javadoc
\definecolor{meucinza}{rgb}{0.5, 0.5, 0.5}
\definecolor{dkgreen}{rgb}{0,0.6,0}
\definecolor{gray}{rgb}{0.5,0.5,0.5}
\definecolor{mauve}{rgb}{0.58,0,0.82}
\definecolor{added}{RGB}{0,128,0}
\definecolor{removed}{RGB}{200,0,0}

\newif\iffinal
\finaltrue   % Uncomment for camera-ready (no comments)
\iffinal

\newcommand{\rev}[1]{#1}
\else

\newcommand{\rev}[1]{\textcolor{red}{#1}}
\fi

\newcommand{\oaw}{\textit{PA}\xspace}%\textsc{OAwPA}\xspace}
\newcommand{\oawo}{\textit{noPA}\xspace}%\textsc{OAwoPA}\xspace}
\newcommand{\mds}{MergeDataset\xspace}
\newcommand{\rds}{RefDataset\xspace}
\newcommand{\mr}{miss-reference\xspace}
\newcommand{\ha}{hybrid analysis\xspace}

\pgfplotsset{compat=1.18}

\AtBeginDocument{%
  }

\setcopyright{acmlicensed}
\copyrightyear{2026}
\acmYear{2026}
\acmDOI{XXXXXXX.XXXXXXX}
\acmConference[]{}{}{}

\acmISBN{978-1-4503-XXXX-X/2018/06}

\begin{document}

% Defina macros para afiliações repetidas
\newcommand{\CInUFPE}{%
  \affiliation{%
    \institution{Universidade Federal de Pernambuco}
    \city{Recife}
    \country{Brazil}
  }
}

\author{Matheus Barbosa}
\CInUFPE
\email{mbo2@cin.ufpe.br}

\author{Paulo Borba}
\CInUFPE
\email{phmb@cin.ufpe.br}

\author{Rodrigo Bonifácio}
\affiliation{%
  \institution{Universidade de Brasília}
  \city{Brasília}
  \country{Brazil}
}
\email{rbonifacio123@gmail.com}

\author{Victor Lira}
\affiliation{%
  \institution{Instituto Federal de Pernambuco}
  \city{Palmares}
  \country{Brazil}
}
\email{vl@cin.ufpe.br}

\author{Galileu Santos}
\CInUFPE
\email{gsj@cin.ufpe.br}

\renewcommand{\shortauthors}{Barbosa et al.}

\title{The Effect of Pointer Analysis on Semantic Conflict Detection}

\begin{abstract}
Current merge tools don't detect semantic conflicts, which occur when changes from different developers are textually integrated but semantically interfere with each other.
Although researchers have proposed static analyses for detecting semantic conflicts, these analyses suffer from significant false positive rates.
To understand whether such false positives could be reduced by using pointer analysis in the implementation of semantic conflict static analyses, we conduct an empirical study.
We implement the same analysis with and without pointer analysis, run them on two datasets, observe how often they differ, and compare their accuracy and computational performance.
Although pointer analysis is known to improve precision in static analysis, we find that its effect on semantic conflict detection can be drastic: we observe a significant reduction in timeouts and false positives, but also a significant increase in false negatives, with prohibitive drops in recall and F1-score.
These results suggest that, in the context of semantic conflict detection, we should explore hybrid analysis techniques, combining aspects of both implementations we compare in our study.
\end{abstract}


\maketitle

\section{Introduction}
 
Textual merge tools operate on a line-by-line basis and report conflicts when developers change the same or adjacent lines of code~\cite{Diff3}. 
More advanced semistructured and structured merge tools~\cite{apel2011semistructured, apel2012structured, cavalcanti2017evaluating, accioly2018understanding, cavalcanti2019impact, 8952450, 10.1145/3360596, da2020detecting} leverage programming language syntax and partial static semantics, and avoid some of the drawbacks of textual tools. 

Textual and structured merge tools, however, are unable to detect \emph{semantic conflicts}, which occur when changes from different developers are textually integrated but semantically interfere with each other~\cite{Horwitz1989IntegratingNV, yang1992program, shao2009sca, brun2013early, barros2017using, sousa2018verified, da2020detecting, zhang2022using}, as when two developers add new assignments to the same variable or state element.
Semantic conflicts often go unnoticed and lead to unexpected undesired behavior~\cite{lima2014abordagem}, as they are hard to detect through typical quality assurance practices like code review and testing~\cite{da2020detecting}.
This way they pose significant risks in both fork-based and centralized projects~\cite{sung2020towards, zhang2022using, mens2002state, Zimmermann, BirdAndZimmermann2012, brun2013early}.
When detected, they are often hard to resolve, especially if discovered late.  

To avoid these problems, researchers have proposed static analyses for detecting semantic conflicts~\cite{barros2017using, binkley1995program, Horwitz1989IntegratingNV,galileu} at merge time, but these analyses often suffer from significant false positive rates.
Part of such false positives are caused by the imprecision typically associated with static analyses, but includes the lack of proper auxiliary Pointer Analysis~\cite{spark, 10.1145/3276511} for implementing conflict detection analyses.
Pointer analysis has the potential to help not only in making method traversal more precise, but also in storing more precise information in analysis data structures; for example, instead of using variable names or types for keeping track of changed state elements, with pointer analysis we rely on the more precise sets of potential references to state elements.

So to understand the effect of pointer analysis on semantic conflict detection, in this paper we conduct an empirical study.
We implement the same semantic conflict detection technique with and without pointer analysis: the former we name \oaw, and the latter we name \oawo.
We run \oaw\ and \oawo\ on a number of merge scenarios (quadruples formed by a base and a merge commit, and its two parents), observe how often the two analyses differ, and compare their accuracy and computational performance.
Our merge scenarios come from two Java datasets.
One of them contains 99 experimental unities manually labeled with semantic conflict ground truth. 
We use this one mainly to assess the accuracy of the analyses (\oaw\ and \oawo), in terms of precision, recall, accuracy, and F1-score.
The other dataset has 907 unlabeled experimental unities, and we use it mainly to understand how often and why the analyses differ, but we also estimate the impact pointer analysis might have on semantic conflict detection accuracy.
We use both datasets for assessing computational performance. 
Our experimental evaluation combines code instrumentation with a more detailed manual analysis.
%\pb{Matheus, no final de tudo, verificar se o que prometemos aqui foi cumprido.}

We find that \oaw\ and \oawo\ differ in less than 14\% of the analyzed scenarios (13.1\% in the small labeled dataset, and 6.8\% in the large unlabeled one).
Although pointer analysis is known to improve precision in static analysis~\cite{10.1145/3276511,10.1145/2499370.2462191, 10.1007/11688839_5}, we observe that its effect on semantic conflict detection can be drastic: \oaw\ brings a significant reduction in timeouts (over 94\% in both datasets) and false positives (44\% in the labeled dataset, and at most 11\% in the other), but also a significant increase in false negatives (28.6\% in the labeled dataset, and at most 10\% in the unlabeled), consequently leading to prohibitive drops in recall and F1-score, when compared with \oawo. 
The \oaw\ reduction in timeouts comes from having to traverse a smaller set of methods, as the use of pointer analysis yields a more precise call graph, avoiding edges to likely unreachable methods.  This, however, can lead to false negatives because pointer analysis is unsound in the presence of reflection, certain kinds of annotations, and other features available in Java.  
The \oaw\ reduction in false positives comes from a more precise call graph, but also from storing more precise information in the analysis data structures.
%\pb{Matheus, confere esses números para ver se eu não falei besteira.}

Contrary to results in other contexts~\cite{10.1145/3276511, 10.1145/3236024.3236041, 10.1145/2499370.2462191, reif2021callgraphs, 10.1145/1925844.1926390, 10.1145/1639949.1640108}, we observe that the precision benefits of pointer analysis hardly compensate its unsoundness drawbacks in the context of semantic conflict detection.
Our results suggest that conflict detection demand novel hybrid analysis techniques, combining aspects of both implementations we compare in our study.
Such a hybrid analysis could use \oaw's ability to reduce timeouts and false positives, while also mitigating its tendency to miss semantic conflicts in the presence of language features that lead to unsoundness. 
We discuss the expected results of this hybrid analysis and provide a comparative assessment based on our current experimental data.

%\pb{Se tiver espaço no final, colocar aqui: The main contributions of this paper are the follwoing: aligned implementations, with and without pointer analysis, of the same static analysis technique for conflict detection; evidence of the effect of pointer analysis on semantic conflict detection; suggestion of a hybrid analysis for conflict detection; replication package...}
%\pb{Se tiver espaço no final, colocar aqui: The rest of this paper is organized as follows...} 

%\rev{Overall, our findings bring several implications into the trade-offs involved in applying pointer analysis to semantic conflict detection.} Our results highlight not only the potential benefits but also the limitations and challenges associated with this technique. We hope that these findings can guide future research efforts toward the development of more robust, hybrid analysis frameworks capable of balancing scalability, precision, and coverage in the context of collaborative software development.

\section{Motivating Example}

To illustrate the effect of pointer analysis on semantic conflict detection, consider the example in Figure~\ref{fig:codigo-motivador}.

\begin{figure}[h]
	\centering
	\begin{lstlisting}[escapechar=/, stepnumber=1, numbers=left  ]
class Text {
  private String t;
  private Report r;
  Text(Report r, String t) {
    this.r = r; this.t = t;
  }
  void generateReport() {
    /\colorbox{red}{\hypertarget{rcountDupWords}{} \textcolor{white}{r.countDupWords();}}/
    r.countComments();
    /\colorbox{blue}{\hypertarget{rcountDupWhiteSpace}{} \textcolor{white}{r.countDupWhiteSpace();}}/
  }
}
  \end{lstlisting}
	\caption{Successful textual merge integrating the changes of two developers, one highlighted in red and the other in blue}
	\Description{Successful textual merge integrating the changes of two developers, one highlighted in red and the other in blue}
	\label{fig:codigo-motivador}
\end{figure}

It shows Java code resulting from a merge that integrates changes independently made by two developers, Left and Right, to a common Base commit.
Left changes appear in red, while Right changes appear in blue.
The remaining code corresponds to the original content in the Base commit.
To improve text report, Left added a call to \texttt{countDupWords}, which counts duplicate consecutive words in the text.
Right independently added a call to the \texttt{countDupWhiteSpace} method, which is responsible to count the number of consecutive whitespace sequences in the text.
As the changes made by Left and Right were not in the same or consecutive lines, textual merge was successful, with no report of a merge conflict.

Now consider a static analysis\footnote{We describe its algorithm in more detail in the following sections.} that detects semantic conflicts whenever the same state element (variable, object field, etc.) is modified by the changes made by two contributors.
Running that for the code in Figure~\ref{fig:codigo-motivador} reports a conflict--- as in Figure~\ref{fig:report-caso-motivador}--- given that the implementations of both \texttt{countDupWords} and \texttt{countDupWhiteSpace} write to the same field; consider \texttt{ReportSimple} in Figure~\ref{fig:codigo-motivador2} to be the only implementation of the \texttt{Report} interface.
In this case we say that \textbf{Right interferes with Left}, as it breaks the change made by Left.
For example, generating a report for the \texttt{"the\textvisiblespace the\textvisiblespace\textvisiblespace dog\textvisiblespace dog"} text would result in \texttt{fixes} being \texttt{1}, ignoring duplicate word counting as the second assignment to \texttt{fixes} overrides the first.

\begin{figure}[h]
	\centering
	\begin{lstlisting}[escapechar=!]
Interference in class Text, method void generateReport(), execution of line 8 overrides 10, assigning to variable this.<ReportSimple: int fixes> 
Caused by line 8 flow:
  at !\textcolor{blue}{\hyperlink{rcountDupWords}{Text.generateReport():8}}!
  at !\textcolor{blue}{\hyperlink{countDupWords}{ReportSimple.countDupWords():4}}!
And line 10 flow:
  at !\textcolor{blue}{\hyperlink{rcountDupWhiteSpace}{Text.generateReport():10}}!
  at !\textcolor{blue}{\hyperlink{countDupWhiteSpace}{ReportSimple.countDupWhiteSpace():9}}!
	\end{lstlisting}
	\caption{Simplified example of an alert reported without the use of pointer analysis}
	\Description{Simplified example of an alert reported without the use of pointer analysis}
	\label{fig:report-caso-motivador}
\end{figure}
Consider now that one provides another implementation of the \texttt{Report} interface, say \texttt{ReportAdvanced} (see Figure~\ref{fig:codigo-motivador2}), which counts duplicated words and whitespaces separately, storing them in different fields.
In such a situation, an analysis that doesn't employ pointer analysis (\oawo) would still report a conflict, as it assumes that both implementations of \texttt{Report} could be invoked at runtime, independently of how \texttt{Text} was initialized.
A pointer analysis (\oaw) based implementation, however, would correctly report a conflict only if it collects information that \texttt{Text} is initialized with a \texttt{ReportSimple} instance; note that the methods in \texttt{ReportAdvanced} don't write to the same state element.
If \texttt{Text} is initialized with a \texttt{ReportAdvanced} instance, \oawo\ incorrectly reports a conflict (a false positive).

\begin{figure}[!h]
	\centering
	\begin{minipage}{0.45\textwidth}
		\centering
		\caption*{ReportAdvanced implementation}
		\begin{lstlisting}[language=Java, escapechar=/, stepnumber=1, numbers=left]
class ReportAdvanced implements Report {
  void countDupWords(){
    ...
    dupWords++; 
    ...                              
  }
  void countDupWhiteSpace(){
    ...
    dupWhiteSpace++; 
    ...
  }
}
    \end{lstlisting}
	\end{minipage}
	\hspace{0.01\textwidth}
	\begin{minipage}{0.45\textwidth}
		\centering
		\caption*{ReportSimple implementation}
		\begin{lstlisting}[language=Java, escapechar=/, stepnumber=1, numbers=left]
class ReportSimple implements Report {
  void /\hypertarget{countDupWords}{} /countDupWords(){
    ...
    fixes++; 
    ...                              
  }
  void /\hypertarget{countDupWhiteSpace}{} /countDupWhiteSpace(){
    ...
    fixes = count; 
    ...
  }
}
    \end{lstlisting}

	\end{minipage}
	\caption{Report interface implementations}
	\Description{Report interface implementations}
	\label{fig:codigo-motivador2}
\end{figure}

This example illustrates how the use of pointer analysis might refine the precision of static analysis for conflict detection.
In this case better precision is achieved because \oaw\ relies on a more precise call graph than the one used by \oawo.
Precision could also be improved by how \oaw\ stores information on the analysis data structures, as discussed later in the paper.
These precision improvements of \oaw, though, come with drawbacks.
Although not illustrated in the example, in the presence of reflection features available in Java \oaw\ might miss actual conflicts when objects are instantiated without using the \texttt{new} command.
These often lead to false negatives.
Because of the involved trade-offs, opting for \oaw\ instead of \oawo\ might not always be the best choice.
So it's important to rigorously assess the benefits and drawbacks of using pointer analysis in the context of semantic conflict detection.

\section{Static Analysis for Semantic Conflict Detection}\label{sec:impl}

The core idea to detect semantic conflicts is to run a number of static analysis in the merged version of the code, which is annotated with metadata indicating instructions modified or added by each developer that contributed to the merge.
The analyses try to explore potential conflicting situations by keeping track of the changes developers make and how they affect \emph{state elements} such as global variables, object fields, and variables that hold method return values. 
 
For assessing the impact of pointer analysis on semantic conflict detection, we focus on the Override Assignment (OA) analysis~\cite{galileu}, which is informally illustrated in the previous section.
OA reports a conflict when both developers contribute to the merged code with changes (additions or modifications) that semantically involve write operations to the same state element, and there is no write operation to that element in the original code (i.e., the common ancestor branch) that intervenes between the the two developers writes in the composed execution. 
Formally, let $L$ and $R$ be branches derived from a common base $B$, and let $w_0 \in L$ and $w_2 \in R$ be write operations to the same state element $s$. 
OA reports a conflict whenever $w_0$ and $w_2$ both write to $s$ and there does not exist a write $w_1 \in B$ to $s$ such that $w_1$ is ordered between $w_0$ and $w_2$ in the resulting merged version of the code, that is, in the integration of changes from $L$ and $R$.
In such a situation, we say that $w_2$--- the change from $R$--- \emph{overrides} $w_0$--- the change from $L$.

For an accurate assessment of the effect of pointer analysis on semantic conflict detection, we implement two carefully aligned versions of OA, one that does not use pointer analysis (\oawo) and another that does (\oaw).
They are built on top of a flexible and extensible architecture that follows the Template Method design pattern to structure variation points on how each version of OA behaves.
The two versions differ significantly in how they analyze allocation sites and construct the call graph used to model method call dependencies.
We use version 4.5.0 of the Soot framework~\cite{vallee2010soot} as the underlying infrastructure for program analysis. 
For performance and practicality, it is important to control the depth and duration of the analyses. 
By default, the analyses explore up to five levels of method calls and is subject to a maximum execution time. 
If this time limit is exceeded, the analysis is interrupted and marked with a timeout status.

The Override Assignment {\bf without} pointer analysis implementation (\oawo) adopts a call graph based on Class Hierarchy Analysis (CHA)~\cite{YannisPA}, which conservatively resolves virtual method calls by assuming that any subclass may implement the method. 
\oawo\ uses the same method as entry point for constructing the call graph and for analyzing the code.   
Since CHA relies on the class hierarchy rather than object instantiations, it does not require locating \texttt{new} instructions. 
This allows us to restrict the call graph to only the relevant parts of the program, reducing its size and improving analysis focus.

When comparing state elements modified by different developers, both analyses consider different types of assignments left-hand sides (LHS). 
We first explain how \oawo\ handles them: 
%
%\todo[inline]{Interessante, n\~{a}o lembrava que a avalia\c c\~{a}o dos elementos de estado eram diferentes, dependendo do uso ou n\~{a}o de PA. Imaginava que o algoritmo para identificar interfer\^{e}ncia era exatamente o mesmo, o que mudava era apenas a constru\c c\~{a}o do CG. Como o algoritmo para identificar interfer\^{e}ncia muda, n\~{a}o podemos tirar conclus\~{o}es baseadas apenas na mudan\c ca do CG. Isso precisa ficar claro nos resultados e discuss\~{o}es.}
%
%\todo[inline]{Muitos detalhes t\'{e}cnicos, o que eh bom. Mas seria legal avaliarmos se todos s\~{a}o necess\'{a}rios para um ICSE paper (10-12 p\'{a}ginas). Talvez fosse interessante incluir alguns exemplos adicionais aqui para compreender OA na sua completude. Eu compreendi, mas tive que gastar um certo tempo. Alguns revisores de ICSE, mesmo da \'{a}rea de an\'{a}lise est\'{a}tica, talvez tenham dificuldade e se beneficiariam de exemplos.}
%
\begin{description}
\item[Local Variables (\texttt{LV})] The analysis checks if the variables belong to the same method and have identical names, since local variables from different methods cannot conflict. It also includes a safeguard against false positives in redundant local assignments that span multiple lines.
\textit{Example:} assignments to \texttt{x} in the same method body conflict. 
  
\item[InstanceFieldRef (\texttt{IFR})] The analysis checks if the types of the target expressions are the same or are related by the subclass hierarchy. 
It also checks whether the fields are identical, and whether the instructions do not belong to the same constructor. 
This last condition avoids false positives in situations where different developers create different instances of objects but use the same constructor.
\textit{Example:} assignments to \texttt{a.x} and \texttt{b.x} conflict only if \texttt{a} and \texttt{b} are of the same class (or compatible subclasses), and the instructions do not belong to the same constructor.

\item[ArrayRef (\texttt{AR})] Similar to \texttt{IFR}, but with array access indexes playing the role of object fields. 
\textit{Example:} assignments to \texttt{arr1[i]} and \texttt{arr2[i]} interfere only if \texttt{arr1} and \texttt{arr2} are of the same class (or compatible subclasses) and the instructions do not belong to the same constructor.
\rev{\item[StaticFieldRef (\texttt{SFR})] The analysis considers two fields equal if they have the same declaring class, are both static, and have the same name and type.} 
%\textit{Example:} assignments to \texttt{Logger.level} in different classes referencing the same static field conflict.}
%\item[StaticFieldRef (\texttt{SFR})] The analysis checks whether the classes and the fields are the same.
%\pb{Antes tinha isto, mas fiquei com dúvida se era isso mesmo, e se precisava comparar isso tudo: The analysis checks whether two references represent the same static field, considering the declaring class, the field name, its type and the confirmation that the field is in fact static.}
%\pb{Cortei o exemplo porque ele não batia com a descrição: por exemplo, a descrição não falava em subclasse, o exemplo fala. Eu acho que não faz sentido comparar subclasse aqui por causa of static field hiding em Java, mas teria que refletir melhor. Se o field não for static, ele poderia estar em um StaticFieldRef? Não responde aqui, mas sim num meet, se houver dúvida ou não for um erro.}
\end{description}
As \oawo\ heavily relies on names and types, and uses CHA, it tends to be significantly conservative. 
As a consequence, it can generate false positives by erroneously classifying assignments as conflicting when, in reality, they involve write operations to state elements in different memory regions. 
%\pb{``Despite some optimizations made automatically by Soot's default settings in the construction phase of Jimple''. Isso aqui é vago e confunde. Teria que explicar melhor qual a configuração. Caso contrário, omitir aqui mas deixar claro no site qual a configuração. Mesmo que vá explicar aqui no texto, esse aqui não é o ponto de explicar isso. Explica antes, lá na parte que vale para as duas análises, perto da explicação de timeout, por exemplo. Aí aqui explica primeiro que \oawo\ é mais conservadora, e só depois explica que a configuração de soot até alivia isso um pouco, mas não muito. Se não for assim, fica muito confuso, embaralha o fluxo do argumento.}

Contrasting with \oawo, the Override Assignment {\bf with} pointer analysis implementation (\oaw) relies on a less conservative call graph based on the Spark points-to analysis framework~\cite{spark, lhotak2003spark}. 
Spark provides more precise (when compared to other alternatives like CHA) modeling of virtual method call dependencies, which is supposed to improve the accuracy of method call resolution.
As Spark needs object creation instructions (\texttt{new}) to correctly identify class instances, \oaw\ adopts different entry points for creating the call graph and for executing the analysis. 
Since objects can be created outside the scope of the analysis entry point, we use all main methods in a project as entry points for the call graph. 
For projects without main methods, like libraries, we conservatively use all public methods as call graph entry points.
%\pb{Tirei aplicações Android porque talvez nelas a gente pudesse identificar algo melhor como entry point. Mesmo para libraries talvez desse para melhorar usando o CHA numa primeira etapa.}

Besides relying on a different call graph, \oaw\ analyzes state elements affected by different developers changes using \textit{points-to} information.
Instead of relying on names and types as explained for \oawo, \oaw\ checks whether the \textit{points-to} sets of the target expressions involved in the assignments overlap.
In the case of \texttt{IFR} and \texttt{AR} instructions changed by different developers, we first check if their target or base array expressions have valid \textit{points-to} sets.
If they don't, we perform a fallback comparison as in \oawo.
Otherwise we report a conflict only if there is a non-empty intersection between the involved \textit{points-to} sets, the accessed fields or array indexes match, and the instructions do not belong to the same constructor.
For instance, in the case of IFR, we report a conflict in case LHS \texttt{a.x} and \texttt{b.x} appear in assignments and both \texttt{a} and \texttt{b} may point to the same allocation site, and the instructions do not belong to the same constructor. 

Both \oaw\ and \oawo\ share a set of common mechanisms that enhance the consistency and effectiveness of the analysis process. 
These include utilities for type checking, conflict tracking, and special handling of constructors. 
The analyses traverse code recursively using the specific call graphs, and apply filtering strategies to ignore irrelevant or unsupported methods. 
In the end, detailed reports and diagnostic logs are generated to support further inspection.

\section{Study Settings}

To investigate the impact of pointer analysis on semantic conflict detection, we perform an empirical study to compare \oawo\ and \oaw, both with respect to analysis accuracy and computational performance.
We pay particular attention to the reduction of false positives (FPs), potential changes in false negatives (FNs), and the frequency in which the two analyses differ.
Guided by this goal, we formulate the following research questions:
\begin{enumerate}[(RQ1)] 
    \item How do \oawo\ and \oaw\ compare in terms of conflict detection accuracy?
    \item What is the magnitude and nature of the discrepancies between \oawo and \oaw?
    \item How do \oawo\ and \oaw\ compare in terms of computational performance? 
\end{enumerate}  
These questions allow us to understand the impact of pointer analysis and provide valuable insights into how to configure the analysis in practice. 
This understanding helps practitioners make informed decisions about trade-offs between accuracy and execution time, ensuring a more effective and efficient analysis. 

\begin{figure}[!h]
  \centering
  \includegraphics[width=1\linewidth]{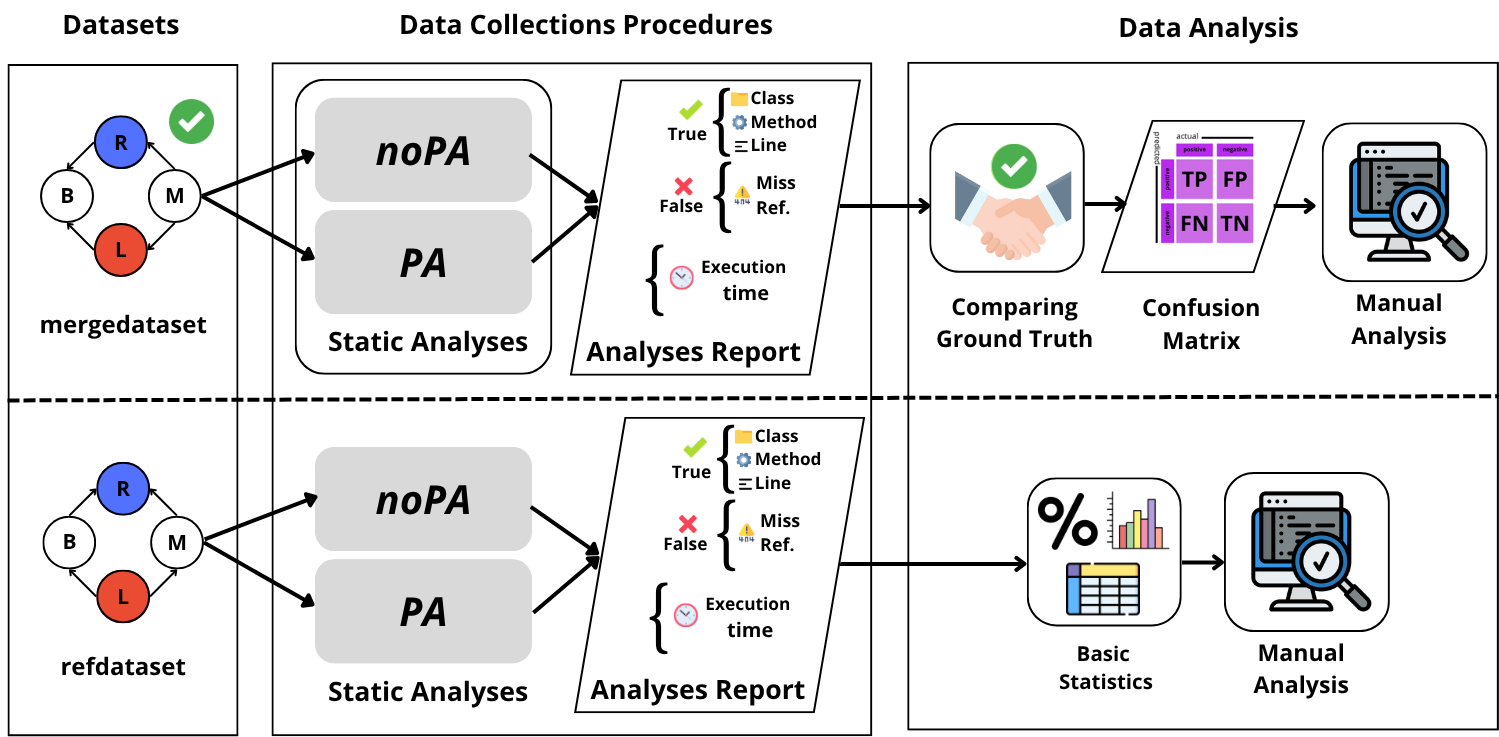}
  \caption{Study settings workflow}
  \Description{Study settings workflow}
  \label{fig:study-settings}
\end{figure}

To answer these research questions, we follow the experimental design illustrated in Figure~\ref{fig:study-settings} and explained in the rest of this section.
%

%\pb{Matheus, verificar se teremos mesmo a terceira questão. Se teremos resultados. Se tiver, teria que adicionar na seção de resultados tb. Verificar se as novas RQs estão ajustadas no resto do artigo.}
\subsection{Datasets}

To answer our research questions, we use merge scenarios from two Java datasets, \mds\ and \rds, which are respectively illustrated in the top and bottom part of Figure~\ref{fig:study-settings}.
The first is labelled with semantic conflict ground truth, which is hard to manually establish, as it often demands deep semantic understanding of the merged code.
The other is unlabeled, but is much larger and diverse, which allows us to better understand the differences between the analyses.

So to answer our first research question (RQ1) and evaluate the accuracy of the proposed analyses (\oawo\ and \oaw) to detect conflicts between contributions from different revisions integrated into a merge scenario, we use the \mds.
It contains merge scenarios in which two contributors changed the same methods or constructors, and comprises a manually labeled set of 99 experimental unities associated with 54 merge scenarios from 32 open-source Java systems. \rev{As a merge scenario might have more than one declaration that was independently changed by both developers, we consider pairs (scenario, declaration) as our experimental units.}
The \mds\footnote{https://github.com/spgroup/mergedataset} scenarios have been used in previous studies on semantic conflicts~\cite{galileu,da2020detecting,sousa2018verified, barros2017using}.

%All scenarios in the dataset were manually analyzed, following a structured double-checking process. According to Jesus et al.~\cite{galileu}, each scenario in the \mds was assigned to two collaborators, who performed individual analyses with the aim of verifying the existence of interference between the modifications of \emph{Left} and \emph{Right} and, thus, establishing the ground truth. Indeed, the ground truth of the \mds used uses the concept of locally observable interference (LOI)---an unintentional interference between changes integrated by different developers. LOIs serve as approximations to the concept of conflict in our research. If the analyses of the two collaborators converged, the result was adopted as the ground truth of the scenario~\cite{galileu}. In case of divergence, the collaborators discussed the scenario to reach a consensus. If consensus was not reached, a third collaborator was involved in the final decision.

%\pb{Achei melhor tirar a parte que fala do processo de estabelecer ground truth. Isso cabe no artigo de galileu, não aqui, quando não seguimos isso. Também tirei a parte de LOI. Tem que trocar isso por conflito no resto do texto. Complica adicionar mais uma sigla, e teria que explicar desde o começo a diferença de conflito e interferência. A definição de LOI que tinha no texto também estava imprecisa.}

To answer our second research question (RQ2) and understand the extent to which the analyses differ when pointer analysis is applied versus when it is not, we additionally use the \rds~\cite{victor}. 
Similarly to the \mds, the \rds contains methods or constructors modified by two contributors. 
The \rds\footnote{\href{https://github.com/victorlira/ref-dataset}{https://github.com/victorlira/ref-dataset}} consists of 907 experimental units rigorously selected and balanced satisfying the following criteria: GitHub's greatest hits repositories with at least 10 stars, using Maven or Gradle for build automation, and with the latest main branch commit successfully compiling and passing tests within 30 minutes on JDK 8, 11, or 17. 
This ensured a higher likelihood of automated \texttt{.jar} file generation, which is needed for running the analyses. 
Lastly, an additional filter was applied to confirm that the target class and method changed by two developers were included in the final build's .jar file. 
This more extensive dataset allows us to better evaluate the differences caused by the analysis of pointers in more varied and less restrictive scenarios where the expected results of the analysis are not previously known.

To answer RQ3, we also use the two datasets.
For each scenario in both datasets, we have the JAR file of the merge commit version without external dependencies, as well as the line numbers of the modifications from \emph{Left} and \emph{Right} in a merge scenario where two developers mutually modified the same method. 
This type of scenario was chosen because it is more susceptible to interference.
%\pb{Cortei as referências para o artigo do SBES~\cite{sbes2022}. Quando o artigo for aceito, a gente coloca.}

\subsection{Data Collections Procedures}

%\todo[inline]{Como aqui deixamos claro o ``depth limit'', podemos retirar tal informa\c c\~{a}o da se\c c\~{a}o anterior. Parce-me mais uma decis\~{a}o de settings do que de implementa\c c\~{a}o.}

With the prepared inputs, we run \oawo\ and \oaw\ for each experimental unit in both datasets. 
For execution time reasons, the analyses execute with a default depth limit of five levels, meaning that they only analyze methods call chains up to five levels deep from the entry point. 
In preliminar experiments, this value has been shown to lead to a good balance between execution time and accuracy of results. 
%\pb{Temos números similares para OA? Os de Galileu acho que são para todas as análises. Se não tiver, melhor tirar isto: ``For instance, Jesus et al. discuss that running the experiment with a depth limit of 10, instead of 5, increases the execution time by a factor of 3 to 6, depending on the analysis~\cite{galileu}. However, a depth limit of 10 does not bring any significant change in the accuracy results.''}  \mb{Não sei, vamos saber melhor com o trabalho de pedro}
The timeout was set to 300 seconds as the maximum time a user should reasonably wait for this type of analysis, while also ensuring the feasibility of running the experiment at scale.

%\todo[inline]{As informa\c c\~{o}es a seguir, ou parte delas, foram discutidas na se\c c\~{a}o anterior. Sugiro remover de l\'{a} e manter apenas aqui.}

We executed our experiment with default settings for the configuration of the Jimple Body Creation (\texttt{jb}) phase of Soot, except for the Use Original Names (\texttt{use-original-names}) option, which preserves the original names of local variables present in the source code.
We activate this setting to make the conflict reports generated by the analyses more readable and informative, so that users can rely only on source code information to interpret them. 
We also configure Soot so that it keeps line numbers from bytecode, and doesn't analyze  classes in standard Java libraries, except for basic classes like \texttt{String}, \texttt{RuntimeException}, and the ones that simply encapsulate primitive types (\texttt{Integer}, \texttt{Long}, etc.). 
So classes like \texttt{ArrayList} and \texttt{HashMap} are discarded from our analysis, for example. 
Just as the builds of the scenarios used in our datasets do not have external dependencies, we also reduce this way the amount of code to be analyzed, especially because effects in no project classes are harder to avoid when trying to solve a semantic conflict.

To avoid variations in software and hardware configuration, we create a \texttt{Docker} file that organizes all the necessary experimental infrastructure. 
This, together with the repositories and detailed instructions for running the experiment, helps repeatability and reproducibility.

%With the environment prepared, we used \href{https://github.com/spgroup/miningframework}{miningframework}\footnote{A framework for mining git projects. Available at: \url{https://github.com/spgroup/miningframework}} as the necessary infrastructure to execute the implementations \emph{OAwPA} and \emph{OAwoPA} in each of the scenarios. 

For evaluation purposes, the result of an analysis is considered positive for a specific experimental unit if the execution reports one or more conflict. 
In addition to collecting the analysis results reporting conflict or its absence, we also gather execution time information for each analysis. 
For controlling for execution time variation, we run each analysis 10 times for each experimental unit.
The analyses were executed on a virtual machine with Linux Ubuntu 22.04.5 LTS (64-bit) operating system, equipped with 2 TB of RAM and an Intel\textsuperscript{\textregistered} Xeon\textsuperscript{\textregistered} Gold 6338 CPU @ 2.00GHz.

\subsection{Data Analysis}

To answer RQ1, we basically compare the execution results yielded by \oawo\ and \oaw\ for  each of the 99 experimental units from the \mds. 
The results are compared against the conflict ground truth provided by the dataset. 
We then compute a confusion matrix and a set of performance metrics, including precision, recall, accuracy, and F1-score, enabling a quantitative assessment of the analyses in relation to the established ground truth. 
%Additionally, we conduct a manual inspection of selected cases for which the analyses yield different results.  This way we are able to understand the impact of pointer analysis on the results.
%\pb{Na dúvida de essa parte final a partir de "Additionally" já não seria parte da resposta de RQ2. Se for, podemos tirar essa parte daqui e colocar um pouco mais adiante.}

To answer RQ2, we analyze data from the execution of our experiment not only with the \mds\ dataset, but also with \rds. 
Unlike \mds, \rds is not labeled, which prevents direct comparison against a ground truth.
However, \rds is significantly larger and more diverse, encompassing a broader range of real-world scenarios. 
This diversity enhances the reliability and generalizability of our findings when analyzing differences between the two analyses. 
To support this comparison, we apply basic descriptive statistical to compute relative frequencies, construct summary tables, and systematically analyze and present the observed discrepancies. 

In addition, we conduct a manual inspection of representative cases to understand the underlying causes of the discrepancies between \oawo\ and \oaw. 
To support this qualitative assessment, we instrumented the analyses implementations with additional logging code. 
This allowed us to extract valuable insights into how pointer analysis influenced the outcomes. 
In particular, our infrastructure includes a logging system that tracks situations in which pointer analysis failed to resolve a reference (classified as a \mr\footnote{This may occur when the points-to set of a variable or reference expression is empty or undefined at the time it is accessed, such as in the case of field or array dereferences. It may also happen when the call graph fails to resolve the target method of a virtual or interface invocation.}).
\label{mr}
This can be strongly suggestive of a conflict that was not reported by \oaw due to Spark unsoundness issues.

We present a table summarizing the results. Additionally, we compute a count of how often such \mr occurred in execution paths that would otherwise lead to conflicts in \oawo. This helps reveal specific patterns in which pointer analysis suppresses potential conflict reports --- not necessarily due to correctness, but due to analysis imprecision.

\rev{Finally, to answer RQ3, we analyze execution time by running each analysis ten times for every experimental unit in both datasets. We then apply the Shapiro-Wilk test to assess normality, followed by the Wilcoxon signed-rank test to evaluate statistical significance. In addition to these tests, we compute descriptive statistics, plotting the distributions, and perform a scenario-by-scenario comparison to determine which analysis performed faster in each case.}

\section{Results}

This section presents the main results of our empirical assessment. 
We begin with the results comparing \oawo and \oaw in terms of conflict detection accuracy. Then, in Section~\ref{sec:rq2-results}, we provide a broader assessment of the changes in the number of reported semantic conflicts, based on both datasets.
Finally, Section~\ref{sec:rq3-results} presents the computational performance results. 
The complete set of results, along with step-by-step instructions for reproducing the experiments, is available in the online appendix~\cite{appendix}.
\subsection{RQ1: How do \oawo\ and \oaw\ compare in terms of conflict detection accuracy?}\label{sec:rq1-results}

Table~\ref{tab:oa-detection-comparison} presents the results obtained
in the detection of override assignment, using the \mds\ and the two implementations we discussed in Section~\ref{sec:impl}. 
%Here we discuss the main observations of these results addressing our first research question.
%
\begin{table}[h]
  \centering
  \caption{Results reported by the analyses using \mds}
  \begin{tabular}{lcccc}
      \hline
      & \textbf{\oawo} & \textbf{\oaw} & \textbf{Diff.} & \textbf{Rel. Change (\%)}    \\
      \hline
      \texttt{true}       & 17  & 10   & -7 & -41.2\%   \\
      \texttt{false}      & 76  & 89 & +13  & +17.1\%    \\
      \texttt{timeout}    & 6   & 0   & -6  & -100.0\%   \\
      \hline
      \textbf{total}      & 99 & 99 \\
      \hline
  \end{tabular}
  \label{tab:oa-detection-comparison}
\end{table}
%
%\str{The analysis with pointers resulted in a 41.2\% reduction in cases classified as \texttt{true} (from 17 to 10) and a 17.1\% increase in \texttt{false} cases (from 76 to 89), compared to the version without pointer analysis. However, this variation does not necessarily imply a change in interference detection, and it is necessary to verify the \textit{ground truth} for validation.}
%
%\str{There was also a complete elimination of \textbf{timeouts} (from 6 to 0), indicating a gain in efficiency. This gain is related to the more precise filtering of the analysis with pointers, which restricts the analysis to only methods of effectively instantiated objects, unlike the CHA-based approach that returns all possible implementations of a method, leading the OA to analyze each one.}
%
%The \oaw analysis resulted in a 41.2\% reduction in cases with semantic conflicts (classified as \texttt{true} in Table~\ref{tab:oa-detection-comparison})---from 17 to 10---and a 17.1\% increase in cases without semantic conflicts (classified as \texttt{false} in the table)---from 76 to 89---compared to the version without pointer analysis (\oawo). However, this variation does not necessarily indicate an improvement in interference detection, as verifying the \textit{ground truth} is necessary for accuracy validation.
%Before turning to the ground truth assessment, 

It's important to note that \oawo\ resulted in six timeout cases in our assessment, whereas \oaw did not lead to any timeouts. 
This indicates a gain in efficiency obtained with \oaw. 
This gain is related to the more precise filtering enabled by pointer analysis, which restricts the analysis to methods of effectively instantiated objects, unlike the CHA based approach that considers all possible implementations of a virtual method, resulting in a more complex call graph, and more methods to traverse.

%\str{To compare with the ground truth and build the table summarizing the both confusion matrices for OAwoPA and \oaw, we chose to keep the 93 scenarios in which both analyses reported true or false, eliminating cases with timeout status. We did this because we wanted to compare only true or false values, and considering timeout as false could bias the results.}

%\todo[inline]{\rb{Poderia colocar os timeouts na tabela?} \mb{Não, porque não consigo saber o resultado esperado dos cenários. Não tenho como identificar se são FP, FN, etc. Para o \oaw, isso é possível, mas isso geraria uma análise desbalanceada, com mais cenários de um lado do que do outro. A menos que eu considerasse os timeouts como falsos, o que você acha disso? Na minha opinião, não seria adequado, pois os timeouts acontecem e possivelmente indicam uma vantagem do uso de análise de ponteiros e esse é, inclusive, um argumento que utilizo ao longo do texto. Mas posso rever se acharem necessário.} \pb{Matheus, lembro da gente ter discutido isso, mas não lembro porque decidimos não considerar os timeouts como falsos. Como ficariam os resultados de acurácia? Ficariam ainda melhores para \oawo? Talvez por isso que a gente optou deixar fora?} \mb{Todos os casos tem LOI false. Os resultados seriam ainda piores para \oaw. Precision 0,20, Recall	0,07, Accuracy 0,65, F1 Score	0,10. A acuracia de \oawo iria para 0.70}}

\subsubsection{Accuracy Assessment}

We exclude from the ground truth assessment the experimental units for which \oawo\ resulted in a timeout--- leading to a total of 93 units from the \mds. 
This way we compare only the cases for which our implementations report either \texttt{true} (indicating a semantic conflict) or \texttt{false} (indicating a conflict-free scenario), as treating timeouts as \texttt{false} could bias the results.
\begin{table}[!h]
  \centering
  \caption{Classification summary for \oawo\ and \oaw}
  \begin{tabular}{lcccc}
      \hline
      & \textbf{\oawo} & \textbf{\oaw} & \textbf{Diff.} & \textbf{Rel. Change (\%)}  \\
      \hline
      \textbf{TP}  & 8  & 2  & -6  & -75.0\% \\
      \textbf{FP} & 9  & 5  & -4  & -44.4\% \\
      \textbf{TN}  & 55 & 59 & +4  & +7.3\%  \\
      \textbf{FN} & 21 & 27 & +6  & +28.6\%  \\
      \hline
      \textbf{Total} & 93 & 93 \\
      \hline
  \end{tabular}
  \label{tab:cfmx-loi}
\end{table}

The results in Table~\ref{tab:cfmx-loi} highlight the impact of using the pointer analysis based implementation (\oaw). 
A significant reduction in \textit{False Positives} (FP) is observed, falling from 9 to 5.
On the other hand, the number of \textit{False Negatives} (FN) increases from 21 to 27. Overall, \oaw\ reduces FPs by 44.4\%, but at the cost of an increase of 28.6\% in FNs. 
So, although \oaw\ is able to reduce the incidence of false positives, it may compromise the sensitivity of the analysis, failing to identify some legitimate conflicts. 
Therefore, the choice of whether or not to use \oaw\ should take this impact into account.

%\todo[inline]{Caso tenha interesse em definir as m\'{e}tricas, melhor colocar tais defini\c c\~{o}es na se\c c\~{a}o settings e fora de um footnote. Isso no ponto onde essas m\'{e}tricas s\~{a}o mencionadas pela primeira vez.}

Table \ref{tab:metrics-loi} presents the results of the accuracy metrics for both analyses.
\begin{table}[!h]
  \centering
  \caption{Accuracy metrics results}
  \begin{tabular}{lcccc}
      \hline
      & \textbf{\oawo} & \textbf{\oaw} & \textbf{Diff.} & \textbf{Rel. Change (\%)}  \\
      \hline
      \textbf{Precision}  & 0.47  & 0.29 & -0.18  & -39.30\%   \\
      \textbf{Recall}     & 0.28  & 0.07 & -0.21  & -75.00\%  \\
      \textbf{Accuracy}   & 0.68  & 0.66 & -0.02  & -3.20\%   \\
      \textbf{F1 Score}   & 0.35  & 0.11 & -0.24  & -68.10\%  \\
      \hline
  \end{tabular}
  \label{tab:metrics-loi}
\end{table}
Both \oawo\ and \oaw\ metrics in the table are low because OA is only one of the analyses used to detect semantic conflicts, that is, many kinds of conflicts cannot be detected by OA, no matter if it relies or not on pointer analysis. 
So, instead of focusing on the absolute values of the metrics, we should focus on the relative metrics changes between \oawo\ and \oaw.
The results indicate that the use of \oaw\ is drastically negative, with significant drops in all metrics except accuracy.
The much lower impact on accuracy is due to the larger number of false negatives in the analyzed sample, and also for the low percentage of TPs in our experiment, as we run only a single conflict detection abalysis: OA.
In summary, the \oaw\ reduction in FPs is far from compensating for the increase in FNs.
%
%Overall, these results highlight a trade-off: \oaw\ increases analysis conservativeness \pb{Errado!} but at the cost of reduced detection capability. 
%According to these results, PA does not show any benefit in any of the analyzed metrics.
Still, we highlight that \oaw\ eliminated all \oawo\ timeouts in our sample. 
Thus, the decision to use pointer analysis should be made considering the application context and the need to minimize timeouts. 
We further discuss this in Section~\ref{sec:Implications}.

\subsection{RQ2: What is the magnitude and nature of the discrepancies between \oawo and \oaw?}\label{sec:rq2-results}

Part of this question is answered with the results of the execution in both datasets.
In particular, in the \mds\ we find that \oaw\ and \oawo\ differ in 13.1\% of the analyzed experimental units (see Table~\ref{tab:oa-detection-comparison}), while in the \rds, they differ in only 6.8\% of the units \rev{These percentages were obtained by dividing the number of differing cases by the total number of cases in each dataset.} Table~\ref{tab:oa-inter-pa-comparison-subset} summarizes the results with the \rds.
\begin{table}[h]
  \centering
  \caption{Results reported by analyses using \rds}
  \begin{tabular}{lcccc}
      \hline
      & \textbf{\oawo} & \textbf{\oaw} & \textbf{Diff.} & \textbf{Rel. Change (\%)} \\
      \hline
      \texttt{true}       & 273 & 243  & -30   & -11.0\%  \\
      \texttt{false}      & 612 & 675  & +63   & +10.3\% \\
      \texttt{timeout}    & 35  & 2    & -33   & -94.3\% \\
      \hline
      \textbf{total}      & 920 & 920 \\
      \hline
  \end{tabular}
  \label{tab:oa-inter-pa-comparison-subset}
\end{table}

We observe here a consistent pattern with the results reported for the \mds\ in Table ~\ref{tab:oa-detection-comparison}.
All results go in the same direction with the use of \oaw, but here, with the larger dataset, the reduction in FPs and increase in FNs are significantly less intense.
Contrasting, the \oaw\ reduction in timeouts with the \rds\ is almost as intense as with the \mds. 

\begin{figure}[h]
  \centering
\begin{tikzpicture}
  % Círculos reduzidos proporcionalmente
  \draw[line width=0.8pt] (0,0) circle (1.5cm);
  \draw[line width=0.8pt] (1.5,0) circle (1.5cm);
  
  % Títulos sobre os círculos
  \node at (0,2.0) {\textbf{\oawo}};
  \node at (1.5,2.0) {\textbf{\oaw}};
  
  % Linhas horizontais de separação
  \draw[line width=0.8pt] (-2.0,0.5) -- (4.5,0.5);   % superior
  \draw[line width=0.8pt] (-2.0,-0.5) -- (4.5,-0.5); % meio
  
  % Labels à direita
  \node at (4.0,0.8) {\texttt{true}};
  \node at (4.0,0) {\texttt{timeout}};
  \node at (4.0,-0.8) {\texttt{false}};
  
  % Valores "true"
  \node at (-0.5,0.8) {39};       % \oawo only (esquerda)
  \node at (0.75,0.8) {234};        % interseção
  \node at (2.,0.8) {9};         % \oaw only (direita)
  
  % Valores "timeout"
  \node at (-0.5,0) {33};
  \node at (0.75,0) {2};
  \node at (2.,0) {0};
  
  % Valores "false"
  \node at (-0.5,-0.8) {0};
  \node at (0.75,-0.8) {612};
  \node at (2,-0.8) {63};
\end{tikzpicture}
  \caption{Intersection of \oawo and \oaw results by classification type}
  \Description{Intersection of \oawo\ and \oaw\ results by classification type.}
  \label{fig:venn-loi}
\end{figure}
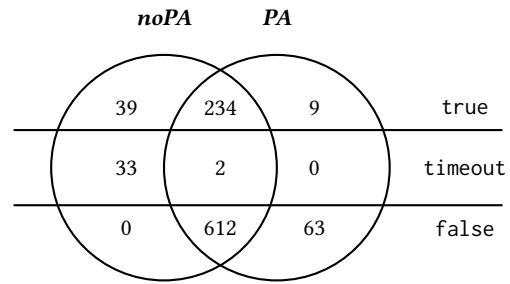

Of the 675 cases classified as \texttt{false} by \oaw, a total of 39 had been reported as \texttt{true} by \oawo. 
This shift is consistent with expectations, as pointer analysis generally improves precision by reducing the number of methods analyzed unnecessarily, likely eliminating false positives. 
A reduction in the number of \texttt{timeout} cases was also confirmed: 33 scenarios classified as \texttt{timeout} in \oawo\ were successfully analyzed by \oaw. 
Specifically, 9 of these were reclassified as \texttt{true}, while 24 were reassigned to the \texttt{false} category. \rev{This value corresponds to the portion of the 33 timeout cases from \oawo\ that are not among the 9 reclassified as \texttt{true} (i.e., 33 - 9 = 24)} The details of these transitions can be visualized in the multi-layered Venn diagram shown in Figure~\ref{fig:venn-loi}. 

%\pb{Como vê esses 24, Matheus? Não está claro. Isso pode gerar confusão. Tem que deixar mais claro de onde os números estão vindo, já que eles não aparecem diretamente na figura/tabela.}

%

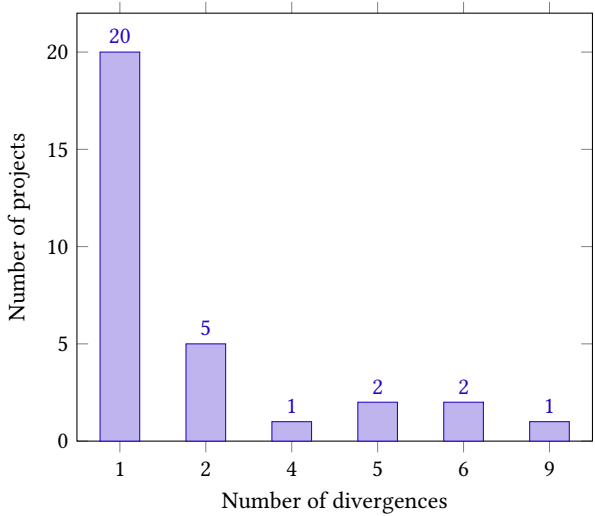
\begin{figure}[h!]
\centering
\begin{adjustbox}{max width=\linewidth}
\begin{tikzpicture}
\begin{axis}[
    ybar,
    bar width=15pt,
    enlarge x limits=0.1,
    xlabel={Number of divergences},
    ylabel={Number of projects},
    xtick=data,
    ymin=0,
    symbolic x coords={1,2,4,5,6,9},
    nodes near coords,
    nodes near coords align={vertical},
]
\addplot coordinates {
    (1,20)
    (2,5)
    (4,1)
    (5,2)
    (6,2)
    (9,1)
};
\end{axis}
\end{tikzpicture}
\end{adjustbox}
\caption{Histogram of the number of divergences per project}
\Description{Histogram of the number of divergences per project}
 \label{fig:histogram-projects}
\end{figure}

\rev{We further investigate how these divergences are distributed among different projects.} Figure~\ref{fig:histogram-projects} presents a histogram of the number of divergences per project. 
The x-axis represents the quantity of experimental units for which \oawo\ and \oaw\ yield different results, while the y-axis indicates the number of projects that exhibited each respective divergence count. 
It can be seen that the vast majority of projects have low values (mostly 1 or 2), with only a few cases slightly deviating, such as \textit{bioformats (9)}, \textit{photon (6)}, and \textit{pebble (6)}. \rev{The complete list of projects can be found in the online appendix~\cite{appendix}.}
This distribution suggests that the results obtained are not biased by a single project or specific subset of projects. 
They are not disproportionately influenced by any individual project.
%\pb{Está ocupando muito espaço e agrega pouco ao que poderia ser descrito no texto de forma resumida. Se precisar de espaço, é um ótimo candidato a ser removido.}

\subsubsection*{Manual analysis of different cases}
To better understand the nature of the discrepancies in these results, we by instrumenting the OA analysis code to report any situation of \mr (see Section~\ref{mr}), which occurs when \oaw\ lacks sufficient information to resolve a \emph{allocation sites} or method target. 
When a \mr occurs, it's possible that the analysis it overlooked a location where the conflict would actually occur. But definitive conclusions can only be made when ground truth is available. 

We begin by analyzing the cases in which \oawo\ correctly reported conflict but \oaw\ didn't, and the ground truth is true. These cases are especially interesting because \oaw, in theory, would increase precision without significantly harming recall.  Understanding such cases can guide improvements in analyzes algorithms and help reduce false negatives.
%These occurrences represent sources of uncertainty in the analysis and typically result from limitations or imprecision in the underlying pointer analysis or call graph construction.
We observed six cases in the \texttt{/mds} project that fall under this condition. In all of them, at least one \mr occurred during \oaw\ execution, which prevented \oaw\ from inspecting the instruction and likely caused it to miss the conflict captured by \oawo. 
Under further manual inspection of these cases, we confirmed that this failure to find potential references was caused by code in which object instances are created through reflection mechanisms, not by directly invoking the Java \texttt{new} command.
These situations directly align with the unsoundness challenges for static analysis of Java reflection reported in the literature~\cite{VinjuSerebrenikSLR,Smaragdakis,Sridharan2013, Kanvar_2016}.

As discussed by Landman et al.~\cite{VinjuSerebrenikSLR}, common challenges for analysis tools include \emph{non-exceptional exceptions}, \emph{programmatic filtering meta-objects}, \emph{semantics of collections}, and \emph{dynamic proxies}. 
\rev{We can observe an example of that in Figure~\ref{fig:retrofit}.}
It shows in red an example of Proxy-based reflection in an experimental unit of the \textit{retrofit} project.
\begin{figure}[!h]
  \centering
  \begin{lstlisting}[language=Java, escapechar=/, backgroundcolor=\color{white}]
    public class RestAdapter {
      public <T> T create(Class<T> service) {
        if (!service.isInterface()) {
          throw new IllegalArgumentException("Only interface endpoint definitions are supported.");
        }
        if (service.getSuperclass() != null) {
          throw new IllegalArgumentException("Interface definitions must not extend other interfaces.");
        }
        /\colorbox{red}{return (T) Proxy.newProxyInstance(service.getClassLoader(),}/
            /\colorbox{red}{new Class<?>[] \{ service \}, new RestHandler());}/
      }      
    }
  \end{lstlisting}
  \caption{Example of an experimental unit from project Retrofit, merge commit \texttt{2b6c719c}}
  \Description{Example of an experimental unit from project Retrofit, merge commit \texttt{2b6c719c}}
  \label{fig:retrofit}
\end{figure}
%
%We highlight in red the statements that use Proxy. According to the documentation\footnote{https://docs.oracle.com/javase/8/docs/api/java/lang/reflect/Proxy.html}, 
\texttt{Proxy} provides static methods for creating dynamic proxy classes and instances, and is also the superclass of all dynamic proxy classes created by these methods. 
In this case, the method dynamically creates an instance of the \texttt{RestAdapter} class. 
This class implements an \texttt{invoke} method that contains the call to \texttt{logAndReplaceRequest}, the entry point method for the OA analyses. 
This explains why \rev{\oaw} fails to report a semantic conflict in this case. 
This is a clear example of computational reflection, which makes it difficult for pointer analysis (Spark in this case) to more precisely infer the objects that a variable can point to.

As previously mentioned, although the \rds is larger than the \mds, it is not labeled. Nevertheless, it is still possible to extract relevant insights from its analysis. We conducted a manual inspection of the 39 experimental units from \rds that fall under the same inspection criteria. This manual analysis allows us to roughly infer which approach was more accurate in identifying conflicts.

Among the 39 scenarios classified as \texttt{false} by \oaw\ but \texttt{true} by \oawo, we observed that 35 contained at least one occurrence of a \mr in the execution analyzed by \oaw, suggesting potential false negatives. A closer inspection revealed that in 25 of these cases, the \mr appeared along paths related to at least one conflict reported by \oawo, supporting the hypothesis that these may be false negatives introduced by \oaw. Conversely, in 14 of the 39 cases, no \mr was found in any path corresponding to the conflicts reported by \oawo, suggesting that \oaw\ may have correctly eliminated false positives and helped reduce noise in the results.

We also analyzed 33 timeout-related divergences in the \rds. Among 24 cases where \oawo\ timed out and \oaw\ returned false, 21 included at least one \mr, suggesting possible false negatives, while the remaining 3 may be true negatives. In 9 cases where \oawo\ timed out and \oaw\ reported true, 6 had associated \mr, indicating potential true positives if unrelated to conflict paths, and the other 3 may also be true positives, as they involved no \mr. But ground truth is needed for confirmation. In contrast, the \mds includes six such cases, all with ground truth labeled as false; among them, \oaw\ reported three as false (true negatives) and three as true (false positives). Log inspection revealed that in both datasets some timeouts in \oawo\ result from CHA inherent inaccuracy, which manifests itself in the generation of over-approximated large call graphs caused by deep class hierarchies that seriously compromise subsequent phases of static analysis and lead the analysis to exceed the execution time limit.

%Overall, these results demonstrate that the analysis of indicators hasn't a negligible impact on the classification of OA cases, affecting only 13.1\% in \mds and 6.8\% in \rds.The findings indicate that \oaw significantly reduced the total number of reported semantic conflict and led to less timeouts cases, compared to \oawo. \pb{acho que isso ai meio que ja foi dito}\mb{sim, mas esse paragrafo deveria ser um resumo geral} A manual inspection of divergent cases suggests that \oaw may introduce false negatives, yet simultaneously eliminates false positives and improves the handling of timeout scenarios. 
%\rb{Essa parte de false negatives eu n\~{a}o estou muito convencido. Sinto falta de evid\^{e}ncia disso.}
%\pb{Tendo a concordar com Rodrigo que essa analise de FPs e FNs pode ser dificil de entender. Ela deveria ser um paragrafo no finalizinho desta secao, resumido, nao com tantos numeros e detalhes. O foco deveria ser mais na natureza das discrepancias. o unico motivo para discrepancias. Teria que comecar top down tb, ta muito bottom up. Fica sem uma narrativa, mais um puro relato do que foi feito.}

%\pb{Cade a discussao sobre a natureza das discrepancias? Ou coloca isso ou ajusta a RQ. O que nao pode é prometer uma coisa e fazer outra.}\mb{isso precisa ser respondido só com analise manual? a anlise dos resultados, o diagrama de ven e o histograma nao respondem isso tbm?}

\subsection{RQ3: How do \oawo\ and \oaw\ compare in terms of computational performance?}\label{sec:rq3-results}

\begin{figure}[h]
\centering
\includegraphics[width=1\linewidth]{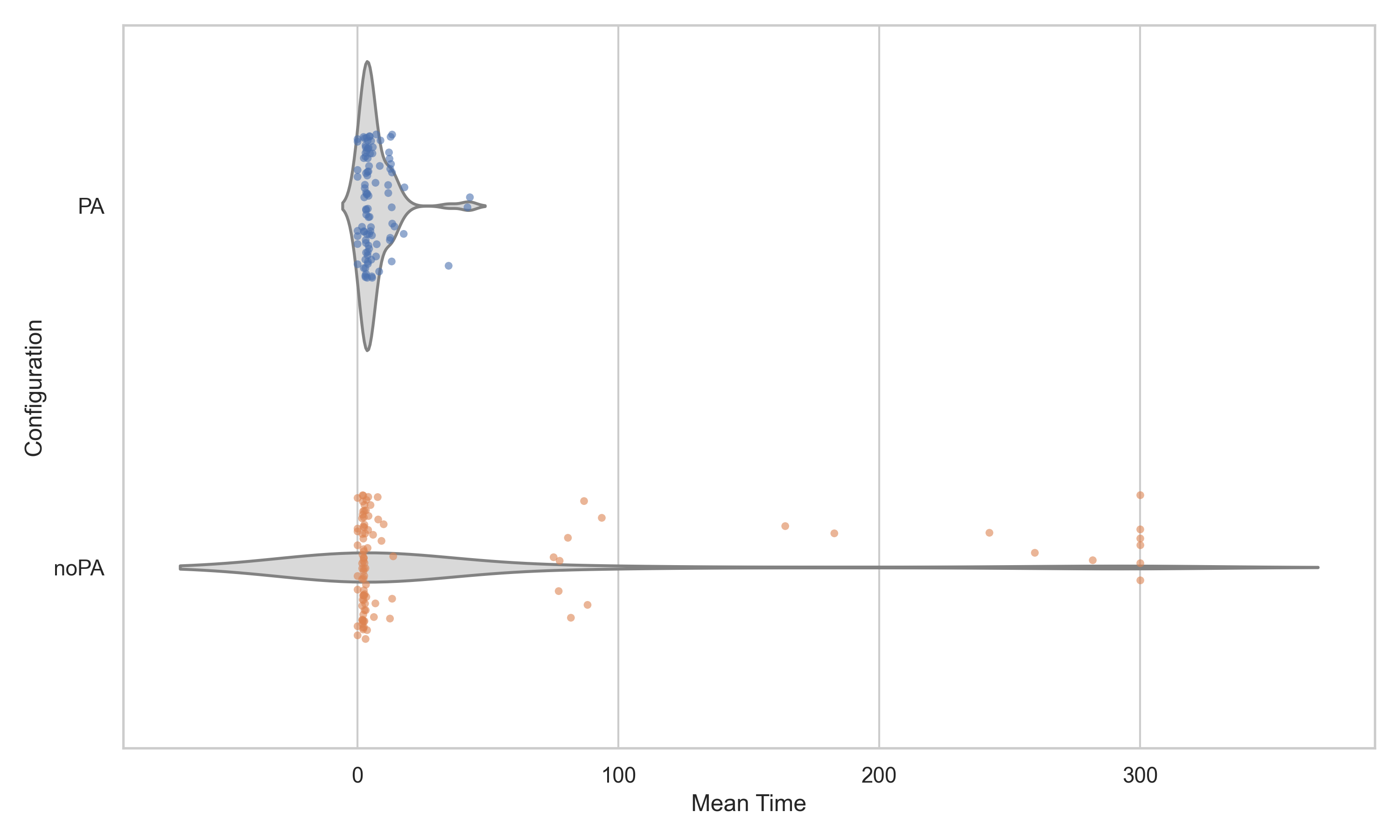}
\caption{Mean execution time per experimental unit in \mds}
\Description{Mean execution time per experimental unit in \mds}
\label{fig:raincloud_plot_mds}
\end{figure}

\begin{figure}[h]
\centering
\includegraphics[width=1\linewidth]{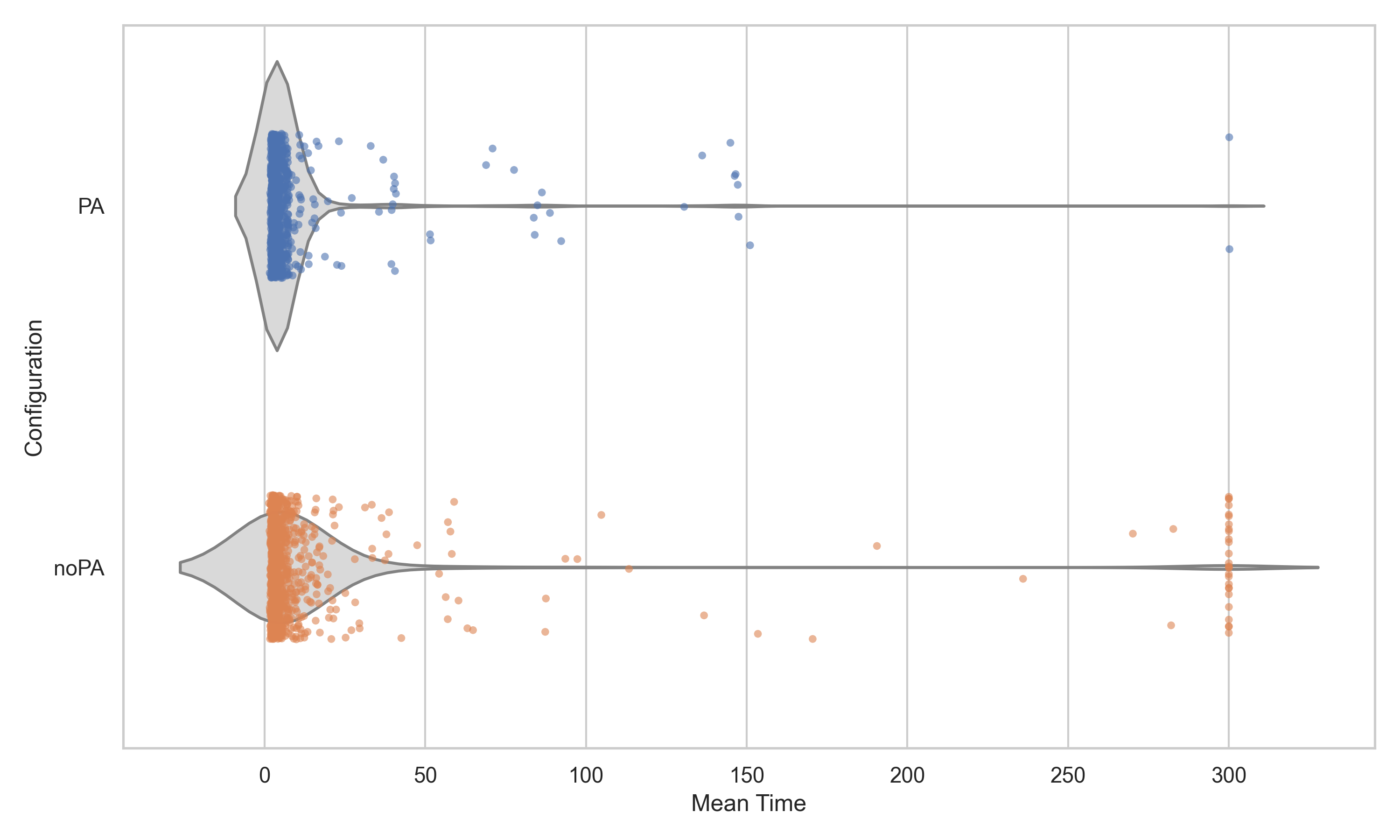}
\caption{Mean execution time per experimental unit in \rds}
\Description{Mean execution time per experimental unit in \rds}
\label{fig:raincloud_plot_rds}
\end{figure}

To assess the computational performance differences between \oawo\ and \oaw, we measure  the average execution time per experimental unit in the 10 executions.  
Figures~\ref{fig:raincloud_plot_mds} and~\ref{fig:raincloud_plot_rds} plot the distribution of mean execution times for the \mds and \rds datasets, respectively.

For \mds, the Shapiro-Wilk normality test indicates that execution times for both configurations significantly deviate from a normal distribution. 
%For \oaw, the test returned W = 0.629 and p-value < 0.001, while for \oawo, W = 0.488 and p-value < 0.001. 
So we applied the Wilcoxon signed-rank test, which resulted in $W = 2326.0$ and a p-value of 0.603, indicating no statistically significant difference between the configurations.

Looking at the data we observe that \oaw was faster in 34 experimental units, whereas \oawo was faster in 65.
Although \oaw\ saves time traversing less methods--- especially in the presence of complex class hierarchies--- \oawo\ saves time with a simpler call graph creation process, and not running pointer analysis.
So faster analysis execution time is highly experimental unit dependent.
On average, \oaw had a total (for all scenarios) mean execution time of 6.47 seconds, while \oawo took considerably longer, with a mean of 38.81 seconds.
This indicates that, considering the whole dataset, the traversal time saving of \oaw, including timeout avoidance, is more efficient. 

%In terms of stability, 24 \oaw scenarios and 18 \oawo scenarios had a standard deviation greater than 10\% of the mean. The highest standard deviation for \oaw was 1.85 seconds (4.4\% of the mean), and the lowest was 0.0 seconds (25.0\%). For \oawo, the maximum deviation was 6.96 seconds (2.9\%) and the minimum was 0.0 seconds (8.1\%). 

We also observe that, in \mds, although \oawo was faster in the majority of experimental units (65), its wins were typically by a small margin—-- often just a few seconds. 
On the other hand, when \oawo lost to \oaw, it tended to lose by a large margin, significantly increasing the global average execution time. 
This explains why \oawo outperforms \oaw more frequently, yet still has a substantially higher average execution time overall. This indicates that a relatively small number of poorly performing cases--- involving complex class hierarchies--- disproportionately affect the overall performance of \oawo.

In contrast, the \rds dataset exhibited higher variability and a more balanced comparison between configurations. The Shapiro-Wilk test again confirmed non-normal distributions. The Wilcoxon signed-rank test revealed a statistically significant difference between the configurations (W = 143341.0, p < 0.001).
A total of 196 \oaw scenarios and 293 \oawo scenarios had standard deviations exceeding 10\% of the mean. For \oaw, the highest standard deviation was 14.42 seconds (9.5\%), while the lowest was 0.01 seconds (0.7\%). For \oawo, the maximum was 16.13 seconds (5.7\%) and the minimum was 0.0 seconds. The overall average execution time was 7.30 seconds for \oaw and 17.26 seconds for \oawo. When comparing scenario by scenario, \oaw was faster in 462 cases, while \oawo outperformed it in 458.

Unlike \mds, the \rds dataset shows a more balanced performance between \oaw and \oawo, both in the number of wins per scenario and in execution time. While \oaw still had a better average runtime overall, the margins between winning and losing cases were smaller and more symmetric than in \mds. The statistically significant result from the Wilcoxon test highlights that this performance difference, though subtle, is consistent. Additionally, the greater number of unstable scenarios (with high relative standard deviation) in both configurations suggests that variability in runtime is a more prominent factor in \rds, potentially influenced by the larger and more heterogeneous set of projects.

In summary, the computational performance comparison between \oawo\ and \oaw\ reveals dataset-dependent dynamics. In the smaller \mds dataset, \oawo often wins more scenarios but suffers from a few costly outliers that significantly increase its average runtime. In the larger and more varied \rds dataset, the performance gap narrows, with \oaw exhibiting a statistically significant but modest advantage in average execution time. This is more consistent with what we should expect in practice.
\subsection{Threats to Validity}
\textit{Construct Validity.}
We assess the effect of pointer analysis for semantic conflict detection with an experiment that includes a single analysis: OA. 
This focus was a deliberate choice, motivated by the need to narrow the scope of the study and enable a precise, in-depth evaluation. 
So our results don't necessarily generalize to other static analyses that are used for conflict detection.
These analyses, however, are similar in nature to OA, relying on call graph information and analysis data structures that could suffer the same kind of effect that was observed in our experiment.
So we don't antecipe analysis characteristics that could lead to different effects from the ones observed here.

\rev{Another threat to the generalizability of our results lies in the choice of call graph construction algorithms. 
We used CHA for the solution without pointer analysis and Spark for the one with pointer analysis. 
Although other algorithms such as RTA could have been explored, our choice was based on practicality and existing evidence. 
Spark is widely regarded as one of the most precise call graph construction algorithms available in the Soot framework, particularly when pointer analysis is enabled, while CHA is a common and efficient option in its absence. 
Even with this limitation, the selected configurations reflect typical usage scenarios and support a meaningful comparison. 
Future studies could explore additional combinations to further assess their impact on conflict detection.}

%\rev{Our study focuses on \emph{locally observable interference} (LOI), considering only the interference that can be identified within the scope of the methods modified by both developers and, recursively, in the methods invoked from them. This choice was motivated by the need to build a reliable ground truth, as limiting the analysis to a local context made manual inspection feasible. Extending the analysis to the entire project would be too complex for manual validation. Although this approach may miss some global interferences, it is a common limitation in previous work and still captures a relevant and practical subset of conflicts.}\pb{Acho que não deveria ter isso. A gente não cria o dataset. E a análise é comparativa, vale para as duas analises. Melhor tb não introduzir o termo LOI aqui, agora.}

\textit{External Validity.}
The datasets used in our study contain scenarios in which different developers concurrently modify the same Java method. 
This particular case of merge scenarios may not represent the entirety of situations encountered in real-world projects, limiting the generalizability of our findings. Projects with more modular code structures, different programming languages, or distinct integration processes may exhibit conflict patterns different from those analyzed.
The size of the \mds\ limits the generalizability of our results, but to the best of our knowledge it's the larger dataset with semantic conflict ground truth, which is considerably hard to establish. 
The \rds\ partly compensate this size limitation, and is the larger dataset we know with buildable merge scenarios, which is necessary for running static analyses.

%\todo[inline]{Tentei revisar o pr\'{o}ximo par\'{a}grafo para ele ser um pouco mais positivo, mas acho que poder\'{i}amos repensa-lo um pouco.}
\textit{Internal Validity.}
\rev{Finally, like any software system, our analyses implementation and experimental infrastructure may contain imperfections despite our best efforts. 
We took great care in the development process and conducted rigorous tests to ensure the correctness and reliability of the code. 
While potential limitations in the implementation of pointer analysis or OA detection could influence the results, we are confident that these do not undermine the overall validity of our findings. 
Nonetheless, ongoing validation and refinement remain important to further strengthen the techniques and their application.}

\section{Implications}
\label{sec:Implications}

Based on the results presented in the previous section, we critically assess the benefits and limitations of employing pointer analysis for detecting semantic merge conflicts.

%\todo[inline]{N\~{a}o sei se dever\'{i}amos sustentar as implica\c c\~{o}es com base em ``qu\~{a}o cr\'{i}ticos s\~{a}o os sistemas''. Isso pode ser um tiro no p\'{e} e podemos discutir sobre isso em uma reuni\~{a}o. Aqui dever\'{i}amos ser mais diretos. Algo como: Os resultados do nosso estudo usando o \mds indicam que \ldots. Por outro lado, os resultados dos experimentos no \rds \ldots. Existem conflitos entre os resultados? Caso afirmativo, por que? Como esses resultados contrastam com o que eh esperado na literatura?}

The results from our study suggest that the effectiveness of pointer analysis varies depending on the system's criticality and specific characteristics under analysis. 
In the \mds dataset, we observed about a 13\% difference in the set of merge scenarios with reported conflicts when pointer analysis (\oaw) was applied, alongside a complete elimination of timeouts and reduction in false positives. 
However, this came at the cost of an increase in false negatives, which were distributed across different projects.
In contrast, the \rds dataset exhibited a smaller difference of only 6.8\% between \oaw and \oawo, the analyses without pointer analysis, indicating that effective conflict detection often relies on code structure and static references, without heavy dependence on pointer analysis.  This is particularly true in systems with limited object relationship complexity, such as those with minimal reflection or well-defined object creation patterns. Additionally, there was a 94\% reduction in timeouts. 
%\rb{o que acontece com bibliotecas?} \mb{acredito que tende a se comportar de forma semelhante}.
Manual inspection also indicated that most of the differences were likely false negatives introduced by the use of pointer analysis. 
%\pb{evitaria repetir aqui os números, o que ja apareceu na secao anterior, e focar mais na discussao mesmo.}

\rev{These outcomes highlight a key trade-off in adopting \oaw in the context of semantic conflict detection with static analysis: it may reduce timeouts and false positives, but at the risk of missing true conflicts, and its overall impact is subtle compared to \oawo
The impact of this trade-off appears to be more significant in datasets with fewer true conflicts (as in \mds), where the removal of even a small number of true positives can greatly affect accuracy. 
This observation aligns with the literature suggesting that pointer analysis can suffer from soundness  issues, especially in the presence of complex object relationships or the heavy use of features such as reflection.} 
%\pb{tá confuso isso aqui, tem que alinhar bem. quando a gente diz precision issues de analise estatica, é porque geram muito FP. mas reflection leva a FN, que é um soundness issue. recall issue.}

%\pb{mesmo problema do uso do termo precision}
While our current implementation does not explicitly handle reflection, there are techniques in the literature, such as reflection-aware static analysis, that attempt to improve accuracy in such cases. However, these techniques often rely on annotations, runtime information, or additional assumptions about the program behavior, which may limit their applicability or introduce new sources of unsoundness. Exploring such techniques as complementary to pointer analysis is a promising direction for future work.

Reflecting on these results, we propose an ideal technique for static analysis-based conflict detection, which we call the \ha . The main idea is to employ an algorithm, such as the one used in \oaw, but switch to the more conservative method of \oawo\ whenever a \mr is encountered. This \ha techniques would avoid introducing additional false negatives while reducing at least the same number of false positives and, at worst, maintaining the same level of timeouts. Based on our data from the \mds, we estimate that this \ha would increase precision by 30.8\% compared to \oawo\ and by 115\% compared to \oaw. Recall would likely match that of \oawo, since all \mr-induced losses in \oaw\ would be avoided. This would provide a more balanced and robust conflict detection strategy.

%\pb{explica um pouco mais porque. e tenta estimar qual seria o up nas metricas, ao inves do drop.} 
%Although we were not able to implement this hybrid approach due to certain technical incompatibilities with the Soot-based infrastructure \pb{melhor omitir isso}, we believe it represents a promising avenue for future work rather than a critical limitation of the current study. 

%\rb{Temos que ter cuidado com isso. Pode dar a impress\~{a}o que eh uma limita\c c\~{a}o severa deste trabalho}. \mb{tentei reformular para suavizar. Por favor verifiquem se podemos melhorar esse techo.}

%Another possibility would be, combining pointer analysis with additional techniques, could avoid a significant portion of the false positives in our sample, such as the refactoring detection tools suggested by \cite{galileu}. \pb{melhor evitar isso}

\section{Related work}
Detecting semantic conflicts in collaborative development is a challenging task, as such conflicts modify program behavior at runtime without triggering syntax or compilation errors. To address this, many studies approximate the problem through \textit{conflict detection}, given the difficulty of inferring developer intent or relying on complete specifications. This section reviews existing work and situates our study as a novel contribution in this area.

The problem of semantic program integration, and the notion of conflict, was formally introduced by Horwitz et al.~\cite{Horwitz1989IntegratingNV}, who proposed an algorithm that uses Program Dependency Graphs (PDGs). Subsequent work, extended this approach to languages with procedure calls, introducing System Dependency Graphs (SDGs)~\cite{horwitz1990interprocedural, binkley1995program}. Tools based on semantic strategies have also been proposed~\cite{336770, yang1992program}, but have long proven impractical. In a more recent work, Barros Filho~\cite{barros2017using} investigated the use of Information Flow Analysis (IFC) with the \emph{JOANA (Java Object-sensitive Analysis) framework}~\cite{joana-paper} tool, which builds SDGs, to estimate conflict in contributions in the same method. Their results indicated information flow in approximately 64\% of the scenarios evaluated, but a manual analysis revealed a high false positive rate (57.14\%) for the existence of conflict. Although theoretically sound, these methods are computationally expensive, taking hours or days for real-world codebases, and often produce false positives due to syntactic changes with no semantic effect. \rev{The static analysis basis, we chose to explore in this work, follows a lighter and more pragmatic strategy by performing analysis only on the merged and annotated version of the code.} By not dealing with four distinct graphs, it avoids the construction of multiple complete SDGs and produces a much faster performance (seconds per scenario). 

In a more recent effort, Sousa et al.~\cite{sousa2018verified} proposed a formal verification approach for three-way merges that verifies that program merges do not introduce new unwanted behaviors, called SafeMerge, which was evaluated on 52 real-world GitHub scenarios. They found that the proposed tool can identify behavioral issues in problematic merges generated by unstructured tools, but with a false positive rate of about 15\% and an average runtime of 0.5 seconds per scenario. Although SafeMerge presents low runtimes, it operates at a different level of abstraction and targets formal verification, while the current work focuses on practical conflict detection with lightweight static analysis.

%Barbosa et al.~\cite{sbes2022} proposed and implemented the \textit{OA} analysis to detect conflict, evaluating it on 78 integration scenarios.

This work builds upon a line of research focused on detecting semantic merge conflicts through lightweight static analysis, centered on the merged version of the code. Santos de Jesus et al.~\cite{galileu} introducing a framework with four interprocedural analyses (Direct Flow, Confluence Flow, Override Assignment, and Program Dependence Graph) applied to the merged code. Their evaluation on 99 real GitHub scenarios showed a median execution time of 17.8 seconds per unit—much faster than test-based or full SDG-based methods. While effective in some cases, the approach suffered from false negatives, suggesting the need to combine it with other techniques. The present study is a direct extension of the OA approach. It investigates the specific role of pointer analysis in enhancing the performance and accuracy of this already lightweight technique. By analyzing pointer analysis impact on call graph construction and its effect on reducing false positives and false negatives, this work deepens the understanding of a previously underexplored factor in the precision of static semantic conflict detection.

Another direction for semantic conflict detection involves test-based methods. Da Silva et al.~\cite{da2020detecting} proposed an approach using automated unit test generation to detect conflict, evaluating it on 38 merge scenarios and identifying 4 true positives and 11 false negatives. Empirical studies by Brun et al.~\cite{brun2013early} and Kasi and Sarma~\cite{Cassandra} also explored the frequency of semantic (test) conflicts, reporting occurrence rates ranging from 3\% to 35\% of merge scenarios. However, such approaches heavily depend on the quality and coverage of the available test suite, which may introduce imprecision and lead to longer analysis times.~\cite{brun2013early, nguyen2015detecting, 10.1145/2568225.2568300}. In contrast, the present work employs static analysis techniques that do not rely on the existence or quality of test cases to identify conflict. While test-based tools may suffer from low recall due to limited coverage, static analysis aims to provide broader and more consistent coverage of program dependencies. 

By discussing the impact of pointer analysis on semantic conflict detection, this work adds to a broader literature that explores the trade-offs between accuracy and efficiency in static program analysis~\cite{10.1145/3276511, 10.1145/3236024.3236041, 10.1145/2499370.2462191, spark, 10.1145/2666356.2594320}. While previous studies focus on pointer analysis to improve overall accuracy (by reducing, for example, the number of casts that can fail or polymorphic calls~\cite{10.1007/11688839_5, 10.1145/3276511,10.1145/3236024.3236041,10.1145/2499370.2462191, 10.1145/1926385.1926390}.) and scalability (by significantly reducing execution times or timeouts~\cite{10.1145/2666356.2594320, 10.1145/3276511,10.1145/3236024.3236041, 10.1145/1639949.1640108}), this study innovates by applying and quantifying these effects in the specific domain of merge conflicts. 

The observation that pointer analysis consistently reduces timeouts is widely corroborated, since approaches such as object-sensitive~\cite{10.1007/11688839_5, popl11b} or scalability-first~\cite{10.1145/3236024.3236041} aim precisely to make analyses more efficient, preventing them from failing due to timeout. However, the increase in false negatives (FNs) and the difficulty of pointer analysis in dealing with reflective behaviors observed in this work are in line with concerns raised by other researchers about unsoundness in call graphs, often attributed to dynamic features of the language, such as reflection and native methods~\cite{reif2021callgraphs,10.1145/3650212.3680333, Kanvar_2016}, and is even an explicit challenge for pointer analysis in benchmarks such as \textit{jython} and \textit{hsqldb}~\cite{10.1145/2499370.2462191,popl11b}. While most work demonstrating improvements in pointer analysis focuses on runtime-related metrics~\cite{10.1145/3276511,10.1145/3236024.3236041, 10.1145/2499370.2462191}, this study offers a unique contribution by explicitly quantifying that, for the reference dataset, pointer analysis changed the classification result in a small percentage of the analyzed cases (6.8\%). Additionally, the use of datasets derived from real GitHub and Maven Central projects differentiates this work from many others that often rely on synthetic benchmarks or consolidated benchmark corpora, such as DaCapo~\cite{10.1145/2499370.2462191, 10.1145/3236024.3236041, popl11b, 10.1145/1639949.1640108, 10.1007/11688839_5}, reinforcing the applicability of its conclusions to real-world scenarios, although generalizations still require caution due to the specificities of the semantic conflicts investigated here.

\section{Conclusions}

Semantic conflict detection in code integration remains a persistent challenge in collaborative software development, as textually compatible changes may semantically interfere, leading to unintended behaviors. Static analyses have been proposed to detect such conflicts, but they often suffer from a high rate of false positives. Our empirical study investigated the impact of pointer analysis on the accuracy and computational performance of such analyses by implementing the same conflict detection technique with and without it. We observed that, while \oaw\ yields a significant reduction in timeouts and false positives, it also a significant increase in false negatives, when compared with \oawo. This trade-off results in a concerning drop in the ability to detect real conflicts.

This unfavorable balance is partly due to the unsoundness of \oaw in the presence of certain language features, such as reflection, where it may fail to resolve allocations sites, thus masking existing conflicts. These findings, highlights the need for innovative and hybrid analysis to semantic conflict detection. Such techniques could combine the reduce timeouts and false positives benefits of \oaw, while also mitigating its tendency to miss semantic conflicts in the presence of language features that lead to unsoundness, paving the way for more robust and reliable merge tools.
% Next, we will describe possible future work in the area, exploring the limitations of this work.

% \subsection{Future work}
% \label{sec:futurework}

% Future research may explore strategies to combine the proposed techniques more effectively. One promising direction is to implement pointer analysis within other forms of analysis, such as direct flow and confluence flow, which were not evaluated in this work but could influence the generalizability of the results to other types of interference detection.

% Additionally, we intend to investigate the most effective way to implement the \textit{Hybrid Approach} proposed herein, aiming to balance precision and conservativeness in conflict detection.

% Ensure at least one citation is present for BibTeX
\nocite{*}

\bibliographystyle{ACM-Reference-Format}
\bibliography{references}

\end{document}